\documentstyle[11pt,iau185,twoside,epsf]{article}

\begin{document}
\title{A Study of Convection with $\delta\,$Scuti Stars in Open Clusters: the
Pleiades} \author{J.-C. Su\'arez$^{1}$,
E. Michel$^{1},~$G. Houdek\altaffilmark{2}, Y. Lebreton$^{1}$, F. P\'erez
Hern\'andez$^{3}$} \affil{$^{1}$Observatoire de Paris-Meudon, DASGAL, UMR
8633,F-92195 Meudon, France} \affil{$^{2}$I.A., University of Cambridge,
Cambridge, U.K.}  \affil{$^{3}$Instituto de Astrof\'{\i}sica de Canarias,
E-38200 La Laguna, Tenerife, Spain}

\begin{abstract}
In this work we propose a preliminary seismic investigation of $\delta$ Scuti
stars in the Pleiades cluster, focusing on potential diagnostics of convection
and core-overshooting. Taking into account the effect of fast rotation in the
modelling, we compare observed frequencies for 4 $\delta$ Scuti stars with
radial linear instability predictions.  A satisfying agreement is reached
between the predicted ranges of unstable modes and those derived from
observations for ``low-mass'' stars ($\sim 1.55 M_{\odot}$). However, a strong
disagreement is found for ``high-mass'' stars ($\sim 1.77 M_{\odot}$), whatever
the mixing length $(\alpha)$ value. These results are compared with previous
ones obtained for Praesepe.

\end{abstract}

 
\section{Introduction}

 Michel et al.\ (1999) compared observed modes for $\delta$ Scuti stars in the
 Praesepe cluster with unstable modes as obtained from a linear
 stability analysis including a nonlocal time-dependent treatment of
 convection. In the present paper, we extend this work to the Pleiades cluster,
 updating several aspects of the procedure. Following Soufi et al.\ (1998) we
 improve the rotation description in the oscillation computations. We take into
 account the effect of fast rotation in the photometric parameters as described
 in P\'erez et al.\ (1999).
\begin{table}[ht]
\caption{Observational data. $N$:~number of frequencies found; $\nu$:~interval of observed
frequencies in $\mu Hz$; $m_V$:~apparent magnitude (Johnson V filter);
$v\sin i$ (km s$^{-1}$):~projected surface linear rotation velocity.} 
\begin{center}
\begin{tabular}{ccccc}
\hline\\[-10pt] Star & $N$ & $\nu (\mu Hz)$ & $m_V$ & $v\sin i$ \\[2pt] \hline\\[-10pt]
\object{HD\,23643} & 4 & [ 197.2, 377.8 ] & 7.76 & 185  \\
\object{HD\,23607} & 7 & [ 216.3, 444.1 ] & 8.26 &  6  \\
\object{HD\,23567} & 7 & [ 253.1, 524.9 ] & 8.30 &  95 \\
\object{HD\,23156} & 7 & [ 242.9, 529.1 ] & 7.51 & 187 

\\[2pt]\hline
\end{tabular}
\end{center}
\end{table}

\section{Observations and modelling}
Observational data for the $\delta$ Scuti stars used in this work are summarized
in Table\,1. All data and references can be found in Fox et al.\ (2001 and these
proceedings).

 The evolution models for the Pleiades cluster have been computed (for a
metallicity of Z=0.014) using the evolution code CESAM (Morel, 1997). We used
OPAL opacity tables (Iglesias \& Rogers, 1996) and the atmosphere is computed
using the Eddington's $T(\tau)$ law. Models with
$(\alpha,d_{ov})=(1.61,0.2),(1.8,0.2)$ and $(1.61,0.1)$ will be called Standard
(Std), $\alpha$- and $d_{ov}$-models respectively. $d_{ov}$ and $\alpha$ are the
overshooting and the mixing-length parameters respectively. An age of $130 \pm
30$\,Myr is found to be representative of this cluster.

\subsection{The correction for the effect of rotation}
 Since we work with fast rotators, a correction for the effect of rotation on
the photomotric parameters is required (Michel et al., 1999). Following P\'erez
et al.\ (1999), we apply such a correction for the selected objects (Fig.\,1).
The correction gives us access to an estimation of masses, radii and rotation
rates, parameters which are used to compute uniformly rotating models.
\begin{figure}[ht]
\begin{center}
\mbox{\epsfxsize=0.7\textwidth\epsfysize=0.5\textwidth\epsfbox{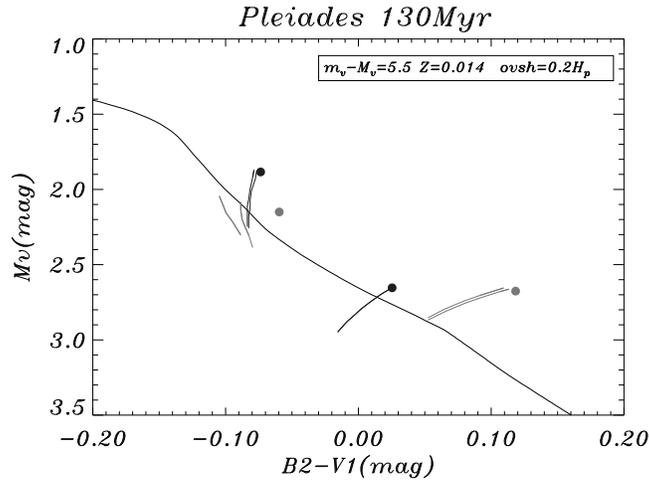}}
\caption{Corrections for the effect of rotation. Circles represent observed
$\delta$ Scuti stars listed in Table\,1. Segments are potential position of
their non-rotating counterparts.}
\end{center}
\end{figure}  
\subsection{The oscillations} 
 
 Rotating evolution models are used to compute adiabatic oscillation
eigenfrequencies. Following Soufi et al.\ (1998), we improve the method used in
Michel et al.\ (1999) to obtain these eigenfrequencies. The objective of such
calculations is to determine a possible range of radial orders associated with
the observed modes given in (Table\,1).
 
 On the other hand, in order to calculate linear growth rates, we use stability 
computations which are carried out in the manner described by Houdek 
et al.\ (1999). To do so, we compute envelope models fitted to our evolution 
models by outer convection zone dephts, masses, radii and effective 
temperatures. 

\section{Results and analysis}
The fundamental parameters obtained in Sect.\,2.1 are not sensitive to realistic
changes in age, $\alpha$ and $d_{ov}$ parameters, as can be expected for such a
young cluster. Thus, envelope models are computed for a unique set of
parameters. The instability computation results will be sensitive to changes in
$\alpha_{nl}$\footnote{The mixing length parameter associated with the non-local
envelope model used in stability computations.}  only through the fine
description of the outer convective regions. They will not be affected by
$d_{ov}$ variations, making useless the study of this parameter here. In this
context, such clusters represent a good opportunity to test the convection in
the outer convective zone, since they are free of uncertainties in the
overshooting description.

\begin{table}
\caption{For {\it Std} and {\it $\alpha$}-models, $n_{ERM}$: predicted radial 
orders from Evolutive Rotating Models\,; $n_{UA}$:~same from Unstability 
Computations\,; $M$:~masses obtained from the correction for rotation.} \tabcolsep=4pt
\begin{center}
\begin{tabular}{cccc}
\hline\\[-10pt] Star & $n_{ERM}$ & $n_{UA}$ & $M/M_{\odot}$\\[2pt] \hline\\[-10pt]
\object{HD\,23643} & [ 1, 7-8 ] & [ 5, 7-8 ] & $\sim 1.77 $ \\
\object{HD\,23607} & [ 1, 7-8 ] & [ 5, 7-8 ] & $\sim 1.77 $\\
\object{HD\,23567} & [ 1, 7-8 ] & [ 1, 6-7 ] & $\sim 1.55 $\\
\object{HD\,23156} & [ 1, 7-8 ] & [ 1, 6-7 ] & $\sim 1.55 $
\\[2pt]\hline
\end{tabular}
\end{center}
\end{table}

The predicted ranges of radial orders and those derived from observations are
given in Table\,2 for {\it Std} and {\it $\alpha$}-models. Contrary to the
Praesepe results, a unique range of radial orders, including the fundamental, is
obtained for the different masses. In the case of Praesepe, a good agreement
between the predicted ranges and those derived from observations was found for
the high-mass\footnote{In the case of Praesepe, where Z=0.019-0.03,
``high-mass'' corresponds to $M>1.8\,M_{\odot}$ and ``low-mass'' to
$M\leq1.8M_{\odot}$.}  stars independently of the $\alpha_{nl}$ values. For
low-mass stars, it was possible to reach an agreement for given $\alpha_{nl}$
values. Here, for the low-mass stars, a good agreement is found, also
independently of $\alpha_{nl}$, however, no matching is possible for high-mass
stars and realistic $\alpha_{nl}$ values.

As in Michel et al.\ (1999), the previous discussion is limited to the results
obtained for instability computations using fundamental parameters of
non-rotating models. For the four $\delta$ Scuti stars of the Pleiades we have
also computed rotating models, which present outer convective zones significantly
deeper than the non-rotating models.  Therefore, for these non-rotating models
it is not possible to build consistent envelope models using the set of
parameters described in section\,2.2 without introducing, somehow, the effect of
the centrifugal force.

\section{Conclusions and perspectives}
  
For selected $\delta$ Scuti stars in the Pleiades, we report here the results of
a comparison between frequency ranges of unstable modes, predicted from
instability computations, with observed ones. Significant differences are
found with respect to results obtained for Praesepe (Michel et al., 1999).  We
find evidence suggesting that these results might be significantly changed when
including some effects of rotation in the stability computations.

Taking advantage of the updated procedure presented in this paper, an
homogeneous simultaneous study of the Pleiades and Praesepe is planned. This
will give us the possibility of determine whether we can obtain consistent
results and how they can constraint the description of the convective transport
in the outer convective zones.

\end{document}